\documentclass[runningheads]{llncs}
\usepackage[T1]{fontenc}
\usepackage{graphicx}
\begin{document}
\title{Do you want to play a game? Learning to play Tic-Tac-Toe in Hypermedia Environments}
%
%
\author{Katharine Beaumont\inst{1}\orcidID{0009-0001-9250-0090} \and
Rem Collier\inst{1}\orcidID{0000-0003-0319-0797} }
\authorrunning{K. Beaumont and R. Collier}
%
\institute{University College Dublin, Dublin, Ireland\\
\email{katharine.beaumont@ucdconnect.ie\\} 
\email{rem.collier@ucd.ie}}
\maketitle              
\begin{abstract}
We demonstrate the integration of Transfer Learning into a hypermedia Multi-Agent System using the Multi-Agent MicroServices (MAMS) architectural style. Agents use RDF knowledge stores to reason over information and apply Reinforcement Learning techniques to learn how to interact with a Tic-Tac-Toe API. Agents form advisor-advisee relationships in order to speed up individual learning and exploit and learn from data on the Web.

\end{abstract}
\section{Introduction}

Agents are fundamental to the earliest formulations of the Web as a network of machine-readable data \cite{berners2001publishing}.
The idea was that these agents could navigate open, discoverable and understandable areas of the Web, interacting with resource in order to achieve their goals. Where agents become aware of one another's presence, this presents opportunities for collaboration and knowledge sharing between them. 

Where is that vision now? It is widely accepted that from a practical perspective, the Semantic Web agent vision has not been progressed \cite{kirrane2021intelligent}. However state-of-the-art research into a new breed of Web agents known as hypermedia agents looks to address this. Hypermedia Multi-Agent Systems (HMAS) is concerned with the creation of MAS that are designed to work with Web architecture \cite{ciortea2019decade}. A key goal of this research is to explore how hypermedia agents can learn and adapt on the Web. 
\cite{vachtsevanou2020long,vachtsevanou2023hyperbrain,vachtsevanou2024enabling} explore this from a single agent perspective using human-led behaviour acquisition, third party LLM models and dynamic machine-readable action descriptions respectively.

This paper illustrates how machine learning techniques can enhance agent learning in a hypermedia multi-agent system\footnote{Code available at \url{https://gitlab.com/mams-ucd/examples/HyperTicTacToe}}. The particular approach we adopt is focused around Transfer Learning (TL), which is typically applied with agents equipped with Reinforcement Learning abilities. TL aims to alleviate some of the traditional shortcomings of Reinforcement Learning, the lack of training data \cite{zhuang2020comprehensive}, or sufficient time to explore and refine the model: in order to learn a policy, an individual agent must take many steps, exploring the environment and potentially performing multiple sub-optimal actions \cite{sutton2018reinforcement}. TL aims to speed up an agent's learning so that it learns faster than it would if it learned individually. Agents with a pre- or partially trained model can transfer some or all knowledge to another learning agent \cite{taylor2009transfer}.

The focus of this demonstration is to illustrate how these concepts and principles can be exploited in a hypermedia MAS using the currently available framework on top of ASTRA\cite{collier2015reflecting}, an agent programming language that is an implementation of AgentSpeak(L) \cite{rao1996agentspeak} and is based on to the Belief, Desire and Intention (BDI) paradigm. This forms part of wider research into how to integrate this type of learning into agent programming in ASTRA and the requirements for language-level changes. 

We present an Application Programming Interface (API) which allows agents to play Tic-Tac-Toe on the Web (TTT API). Multiple agents can interact with the API and each other. All are deployed on the Web and interact using HTTP protocols\footnote{https://www.w3.org/Protocols/}. Agents use open actions: they navigate links and forms, without hard-coded knowledge of the endpoints. This reflects a move towards general models of interaction with the environment from bespoke actions to more abstract, open actions constrained by environmental signifiers such as hypermedia form actions and HTTP verbs.

Tic-Tac-Toe is a well understood agent learning problem that has a small, finite state set \footnote{$3^{9}$, compared to Backgammon with over $10^{20}$ \cite{tesauro1995temporal}} but it cannot be solved through classical techniques as they assume a certain opponent strategy \cite{sutton2018reinforcement}. It is used here to demonstrate how hypermedia agents can learn and engage in Transfer Learning in a Web environment.

\section{Framework Overview}

\begin{figure}[h]
  \centering
  \includegraphics[width=0.95\linewidth]{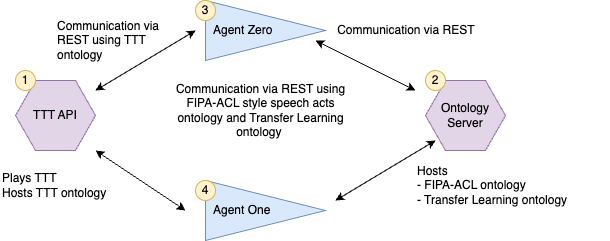}
  \caption{System Diagram}
  \label{fig:sysdiag}
\end{figure}

\subsection{Environment: Tic-Tac-Toe API}

The environment in this demo is the TTT API (1 in Figure~\ref{fig:sysdiag}), a bespoke REST (REpresentational State Transfer) \cite{fielding2002principled} API that includes hypermedia controls such as hyperlinks (links) and forms.  The TTT API returns JSON-LD\footnote{https://json-ld.org/} representations which include links to the ontologies used, allowing agents to reason about and learn from the information presented, and share domain knowledge with other agents.

The ontologies used in the TTT API include the Hypermedia Controls Ontology (in the below example, this is the default or \verb|"@vocab"|)\footnote{\url{https://www.w3.org/2019/wot/hypermedia\#}}, the Hypertext Transfer Protocol (HTTP) Ontology (indicated by \verb|"href"| in the example below)\footnote{\url{http://www.w3.org/2011/http\#}}, the Web of Things (WoT) Thing Description 1.1 (\verb|"wot"|) \footnote{\url{https://w3c.github.io/wot-thing-description/}} as well as the Tic-Tac-Toe Ontology (\verb|"ttt"|). The Tic-Tac-Toe Ontology was created to provide the concept of a game, moves, players, player roles, squares, the board and the result.

An Ontology Server (2 in Figure~\ref{fig:sysdiag}) hosts two additional custom ontologies written to provide FIPA-ACL style communication and basic Transfer Learning concepts. 

The TTT API has a single entry point from which agents can navigate the system (e.g. \verb|http://ttt.api/|)\footnote{All URLs are simplified for readability}. From this entry point, agents can register to play and receive an assigned game identifier (ID) (for example \verb|id1234|), view the board, make moves, and view the result when the game has been completed. The JSON-LD representations combine data with links and forms that specify contextually valid interactions. For example, the response to a successfully registering includes links back to the entry point, the game board, and forms for each of the possible game play moves:

\begin{verbatim}
"links": [
{
  "href": "http://ttt.api/",
  "htv:methodName": "GET"
},
{
  "href": "http://ttt.api/Board?id=id1234",
  "htv:methodName": "GET"
}],
"forms": [
{
  "href": "http://ttt.api/Square11?id=id1234",
  "contentType": "application/json",
  "htv:methodName": "PUT",
  "wot:op": "writeproperty"
},...]
\end{verbatim}

A TTT API game bot executes the game play strategy against the agent. When a new agent registers to play, a new RDF graph of the instantiated ontology classes is created and stored in memory. When moves are made, an instance of the Move class is created along with relationships which link the Move to a Square, \verb|inSquare|, and with a player role, \verb|moveTakenBy|. On querying the result, a JSON-LD representation is returned, and in addition the RDF graph\footnote{https://www.w3.org/TR/rdf12-concepts/} of the game is written out to a file, which provides the possibility to analyse past games and construct a knowledge graph of the game history.

\subsection{Individual Agent Learning}

\begin{figure}[h]
  \centering
  \includegraphics[width=0.95\linewidth]{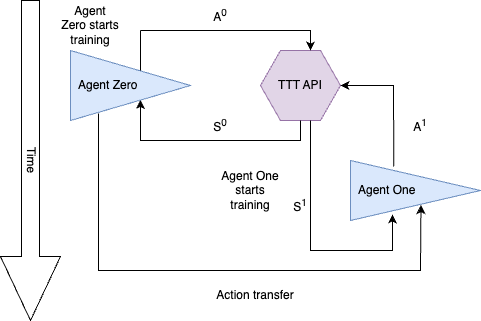}
  \caption{The agent learning process}
  \label{fig:agentlearn}
\end{figure}

The game playing agents (3 and 4 in Figure~\ref{fig:sysdiag}) are implemented using  Multi-Agent Microservices (MAMS), an architectural style and associated framework for deploying multi-agent systems within a microservices architecture \cite{w2019mams}. The framework is implemented on top of ASTRA.

Figure \ref{fig:agentlearn} demonstrates Agent Zero interacting with the TTT API over successive training episodes (complete games) in order to start forming a policy for game play. 

Agents use the module API\cite{collier2015reflecting} of ASTRA in order to interact with the learning mechanism (the \verb|Policy| module, discussed below), and an Apache Jena module (\verb|KnowledgeStore|) as per \cite{beaumont2021collaborative,o2021building}. The \verb|KnowledgeStore| module allows the agents to individually interact with the TTT API. As well as communicating with the TTT API, this module facilitates the creation, parsing and storing of triples and acts as an information store for the game play. Once a response is received from the TTT API, the \verb|KnowledgeStore| fires an event to the agent which triggers a plan to extract information about game play in the agent code. For example, in processing the example JSON-LD response from the previous section, the belief 

\verb|+form_actions(string, string, list)| 

may be added with the details : \verb|"http://ttt.api/Square11?id=id1234"|, \verb|"PUT"|, \verb|["@id"]|. The agent can react to the addition of beliefs regarding forms and use the endpoint, HTTP method, and required input details to perform actions without hardcoded prior knowledge.

Support for the use of Reinforcement Learning (RL) is realised through the custom \verb|Policy| module, which implements an abstract framework that can support multiple RL algorithms.
In this example, each agent uses the Q Learning algorithm \cite{sutton2018reinforcement} with different state/action representations and optimisations. Agents infer a positive reward at the termination of a game if they win, a negative reward for loss and 0 for a draw. 

Agents have a second, separate knowledge store for inter-agent communication using the custom FIPA-ACL style ontology, which in future implementations would allow agents to reason about their communications

\subsection{Transfer Learning}

In Figure \ref{fig:agentlearn}, after Agent Zero starts interacting with the TTT API, Agent One subsequently begins training in the same manner. However in addition to selecting an action based on it's policy during game play, it can also query Agent Zero for a recommended action. The agents have an ad hoc advisor-advisee relationship \cite{da2017simultaneously}, and a limit is set on the number of times agents can query each other to reduce the interruptions to individual learning in the advising agents.

The protocol to request and receive (or not receive) advice is executed over successive messages via HTTP between the agents. An example of the body of an interaction between Agent One and Agent Zero is given below:

\begin{verbatim}
"fipa:request": { 
    "fipa:sender": "http://agent.one", 
    "fipa:receiver": "http://agent.zero", 
    "fipa:reply-to": "http://agent.one/response",
    "fipa:conversation-id": "msgid123",
    "fipa:ontology": "http://ontology.server/transfer#",
    "fipa:content": {
        "http://agent.zero": {
            "tf:Query": "tf:Action",
            "tf:hasState": "http://ttt.api/Board?id=id1234"
           
        }
    }
}
\end{verbatim}

The FIPA-ACL ontology is prefixed with \verb|"fipa"| and the Transfer Learning ontology, with \verb|"tf"|. This demonstrates Agent One requesting the action (\verb|tf:Action|) for a specific board state (\verb|http://ttt.api/Board?id=id1234|). 

Agent Zero either sends a failure message or a recommended action. The recommended action includes a score. This is the value associated with the action, normalised across all of the existing values in the agent's policy, and indicates a measure of confidence in the action. Agent One can reject the advice if the score falls below a threshold. 

The agents are programmed to wait for responses from each other and from the API to account for network latency, and re-attempt communication or continue training if the wait for a response is over a threshold, ensuring fault tolerance.

The effectiveness of Transfer Learning is evaluated by comparing metrics including the total reward gained during single agent learning versus an agent learning engaged in Transfer Learning (the advisee), over a set number of training episodes. Averaged results over multiple evaluation runs show that the advisee has higher discounted and undiscounted reward, and shorter time and number of episodes to a reward threshold, compared to single agent learning, demonstrating accelerated learning.

\section{Discussion}

The demonstration utilises open actions in order to allow agents to react to what may be a changing API. This assumes that the environment conforms the constraint that the responses only include valid interactions and follow HATEOAS principles. 

Learning is integrated into agents in a modular manner, and in this sense the integration with learning is soft as per the discussions in Erduran (2022) \cite{erduran2022machine}: the approach demonstrated provides some controls (agents can start, pause or stop the learning process via actions on the \verb|Policy| module), but as a modular integration it does not take advantage of the flexibility of the BDI agent paradigm. 

This demonstration provides a starting point for the analysis of actions and behaviours that can be extracted and generalised to provide more general mechanisms for learning in hypermedia environments, with a view to language-level supports for learning in agents programmed in ASTRA.

\begin{credits}
\subsubsection{\ackname}CONSUS is funded under Science Foundation Ireland's Strategic Partnerships Programme (16/SPP/3296) and is co-funded by Origin Enterprises Plc.

\end{credits}
%
%
%
\bibliographystyle{splncs04}
\bibliography{bibliography}

\end{document}